\documentclass[twocolumn,aps,showpacs,preprintnumbers,lettersize]{revtex4}
\usepackage{amsfonts}
\usepackage{amsmath}
\usepackage{amssymb}
\usepackage{graphicx}

\begin{document}

\title{Potential-energy (BCS) to kinetic-energy (BEC)-driven pairing in the
attractive Hubbard model}
\author{B. Kyung$^{1}$, A. Georges$^{2}$, and  A. -M. S. Tremblay$^{1}$}
\affiliation{$^{1}$D\'{e}partement de physique and Regroupement
          qu\'{e}b\'{e}cois sur les mat\'{e}riaux de pointe, Universit\'{e}
          de Sherbrooke, Sherbrooke, Qu\'{e}bec, J1K 2R1, Canada  \\
             $^{2}$Centre de Physique Th\'{e}orique, \'{E}cole Polytechnique,
          91128 Palaiseau Cedex, France}
\date{\today }

\begin{abstract}
The BCS-BEC crossover within the two-dimensional attractive Hubbard model
is studied by using the Cellular Dynamical Mean-Field Theory
both in the normal and superconducting ground states.
Short-range spatial correlations incorporated in this theory remove the normal-state quasiparticle peak and
the first-order transition found in the Dynamical Mean-Field Theory,
rendering the normal state crossover
smooth.
For $U$ smaller than the bandwidth, pairing is driven by the potential
energy, while in the opposite case it is driven by the kinetic energy, resembling
a recent optical conductivity experiment in cuprates.
Phase coherence leads to the appearance of a collective Bogoliubov mode
in the density-density correlation function and to the sharpening of the
spectral function.
\end{abstract}

\maketitle

%%%%%%%%%%%%%%%%%%%%%%%%%%%%%%%%%%%%%%%%%%%%%%%%%%%%%%%%%%%%%%%%%%%%%%%%%%%%

   The problem of the crossover between the BCS and the Bose-Einstein
condensation (BEC)
has been of great interest in the context of the pseudogap observed
in underdoped cuprates~\cite{TS:1999}.
The recent discovery of the BCS-BEC crossover in ultracold fermionic atoms
trapped in optical lattices~\cite{OPTICAL_LATTICE}
has renewed our interest in this issue.
%Does the condensation of loosely bound Cooper pairs at weak coupling and
%the BEC of tightly bound composite pairs at strong coupling
%evolve smoothly one into the other
%as the attraction between fermions is changed gradually?
Does the condensation of loosely bound Cooper pairs
evolve smoothly into the BEC of tightly bound composite bosons
as the attraction between fermions is increased gradually?
This question was first addressed by Leggett~\cite{Leggett:1980}
who proposed that these two pictures are limiting cases of a more
general theory in which both the fermionic nature of individual particles
and the bosonic nature of pairs must be considered on an equal footing.
By using a $T-$matrix approximation in the intermediate coupling regime,
P. Nozi\`{e}res and Schmitt-Rink~\cite{NS:1985} extended Leggett's
analysis to a lattice and to finite temperature to find a smooth
connection between the two limits.
Since then, various other approximate schemes
have been used to understand
the pseudogap phenomena as well as the BCS-BEC crossover.
For example, Quantum Monte Carlo (QMC)
simulations~\cite{RTMS:1992,SPSBM:1996,Singer:1999}
of the two- and three-dimensional
attractive Hubbard model found a breakdown of Fermi liquid
theory in the normal state, accompanied by a pseudogap and a spin gap.

   The first Dynamical Mean-Field Theory (DMFT)~\cite{GKKR:1996} study of the BCS-BEC crossover at arbitrary attractive
interactions was carried out by Keller {\it et al.}~\cite{KMS:2001}.
The authors calculated the transition temperature for superconductivity,
which smoothly interpolates from the BCS behavior at weak coupling
to the $t^{2}/U$ behavior at strong coupling.
However, the double occupancy,
the spin susceptibility, and the quasiparticle weight
show, in the $T \rightarrow 0$ limit of their {\it normal state}
solution, a discontinuity
near $U \sim 1.5W$ with $W$ the bandwidth.
Capone {\it et al.}~\cite{CCG:2002} studied the first-order transition
in detail by using exact diagonalization as the impurity solver for DMFT
and later extended their work to finite temperature~\cite{TBCC:2005}.
The discontinuity found in the normal state of DMFT
may suggest a {\it radically} different mechanism for superconductivity
at weak and strong coupling,
%despite the fact that
but that is problematic in the context of the expected
smooth BCS-BEC crossover (in the superconducting state).
Nevertheless, in a recent optical conductivity experiment in the cuprates,
Deutscher {\it et al.}~\cite{DSB:2005} found that near optimal doping there is a reversal
of the sign of the kinetic energy difference between the superconducting (SC) and normal (NR)
states. Although the mechanism for superconductivity in the cuprates
is still under debate, this intriguing experiment calls for explanation
also in the context of the BCS-BEC crossover, as the authors noted.

In the BCS limit, it is well established that superconductivity is potential-energy driven
since the broadening of the Fermi surface caused by superconductivity
leads to an increase in kinetic energy in the superconducting state. At strong coupling,
in the BEC limit, one may expect that superconductivity is kinetic-energy driven
based on two different arguments : a) The gain in potential energy occurs at high temperature where the bosons form.
At low temperature, to leading order in a low density expansion~\cite{Randeria}, the bosons condense for
the same reason as free bosons, because of the gain in kinetic energy. Hence, comparing
with a normal state where the bosons are formed, the superconductivity occurs because
of gain in kinetic energy. b) One can map the attractive Hubbard model
to the half-filled repulsive model in the presence of a magnetic field. The antiferromagnetic order that appears
close to half-filling in this model is analogous to superconductivity and at strong coupling, it occurs because having
neighbors that are ordered leads to a gain in exchange energy~\cite{Millis}. One can check that in the mapping
of the Hubbard model to the Heisenberg or t-J model, the exchange energy corresponds to minus twice the potential
energy of the original Hubbard model~\cite{Fazekas}. These physical arguments do not tell us whether
the change in pairing mechanism occurs in a continuous or discontinuous manner, at what
coupling they occur or what is the order of magnitude of the condensation energy. The
explicit calculations presented in the  present paper thus answer both qualitative and quantitative
questions.

   First we emphasize our main results:
1) Including short-range spatial correlations explicitly removes
   the first-order transition found in DMFT,
   rendering the crossover in the NR state
   also smooth, without a discontinuity, just as in the SC state.
2) Near $U$ equal to the bandwidth of $8t$, a change in pairing mechanism occurs.
   For $U < 8t$ the condensation energy is lowered
   by the potential energy while for $U > 8t$ it is lowered by the kinetic
   energy, resembling a recent optical conductivity
   experiment~\cite{DSB:2005}.
3) The phase coherence manifests itself most dramatically by the appearance
   of a collective Bogoliubov mode in the density-density correlation
   function and by the sharpening of the spectral function compared with the normal state.

   Using Cellular Dynamical Mean-Field Theory (CDMFT)~\cite{KSPB:2001}, we study the crossover between weak and strong-coupling in the two-dimensional attractive Hubbard model~\cite{MRR:1990}
\begin{eqnarray}
H = \sum_{\langle i j \rangle ,\sigma }t_{i j }
      c_{i \sigma}^{\dagger }c_{j \sigma }
    - U \sum_{i}n_{i \uparrow }n_{i \downarrow }
    - \mu \sum_{i \sigma} c_{i \sigma}^{\dagger }c_{i \sigma },
                           \label{eq10}
\end{eqnarray}
where $c^{\dagger}_{i\sigma}$ ($c_{i\sigma}$) are creation
(annihilation) operators for electrons of spin $\sigma$,
$n_{i\sigma} = c^{\dagger}_{i\sigma}c_{i\sigma}$ is the density of
$\sigma$ spin electrons, $t_{ij}$ is the hopping amplitude equal to $-t$
for nearest neighbors only,
$-U$ is the on-site attractive interaction with $U > 0$ and
$\mu$ is the chemical potential controlling the electron density.
The CDMFT method is a natural generalization of the single site
DMFT~\cite{GKKR:1996} that incorporates short-range spatial correlations.
In the CDMFT construction~\cite{KSPB:2001,BK:2002} the infinite lattice is
tiled with identical clusters of size $N_{c}$,
and the degrees of freedom in the cluster are treated exactly
while the remaining ones are replaced by a bath
of non-interacting electrons that is determined self-consistently.
Since the CDMFT method treats short-range spatial correlations explicitly,
it is able to describe features caused by finite dimensionality or
finite coordination number,
which are missed in the single site DMFT.
\begin{figure}[tbp]
\includegraphics[width=8.0cm]{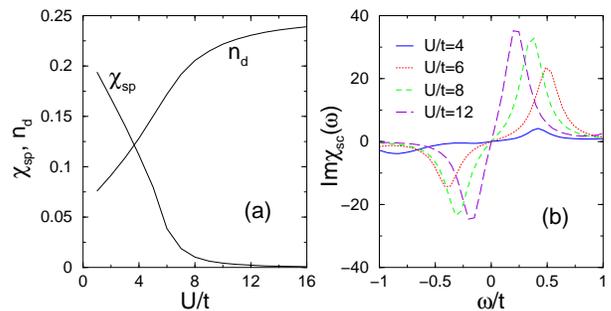}
\caption{(a) Uniform static cluster spin susceptibility $\chi_{sp}$ and double
         occupancy $n_{d}=<n_{i\uparrow}n_{i\downarrow}>$.
         (b) Imaginary part of the $s-$wave cluster pair correlation function
         with total momentum equal to zero $\chi_{sc}(\omega)$.
         The above quantities are calculated in the NR ground state
         at quarter filling ($n=1/2$).
        }
\label{spin_dblocc_sc.fig}
\end{figure}

   To solve the quantum cluster embedded in an effective SC
medium, we consider a cluster-bath Hamiltonian of the form~\cite{Kyung:2005,Kancharla:2005}
\begin{eqnarray}
H && = \sum_{\langle \mu \nu \rangle ,\sigma }t_{\mu \nu }c_{\mu \sigma
}^{\dagger }c_{\nu \sigma }-U\sum_{\mu }n_{\mu \uparrow }n_{\mu \downarrow }
\nonumber \\
+ && \sum_{m,\sigma,\alpha }\varepsilon^{\alpha}_{m\sigma }a_{m\sigma }^{\dagger \alpha }a_{m\sigma}^{\alpha}
+ \sum_{m,\mu ,\sigma,\alpha }V^{\alpha}_{m\mu \sigma }( a_{m\sigma }^{\dagger \alpha}c_{\mu \sigma}
+ \mathrm{H.c.} )
\nonumber \\
+ && \sum_{m,\alpha}\Delta( a_{m\uparrow }^{\alpha}a_{m\downarrow }^{\alpha}
                     + \mathrm{H.c.} ) \, .  \label{eq20}
\end{eqnarray}%
Here the indices $\mu ,\nu =1,\cdots ,N_{c}$ label sites within the cluster,
and $c_{\mu \sigma }$ and $a_{m\sigma }^{\alpha}$ annihilate electrons
on the cluster and the bath, respectively. In the present study we used $N_{c}=4$ sites for the cluster
(minimum number of sites reflecting the full square lattice symmetry) and
$N_{b}=8$ sites for the bath with $m=1,...,4, \alpha=1,2$.
$t_{\mu \nu }$ is the hopping matrix within the cluster and, using symmetry,
$\varepsilon^{\alpha}_{m\sigma }=\varepsilon^{\alpha}$ is the bath energy and $V^{\alpha}_{m\mu \sigma }=V^{\alpha}\delta_{m,\mu}$
is the bath-cluster hybridization matrix.
$\Delta$ represents the amplitude of $s-$wave SC
correlations in the bath.
To deal with superconductivity, the Nambu spinor representation
is used for the cluster operators
\[ \Psi^{\dagger} = ( c_{1 \uparrow}^{\dagger },c_{2 \uparrow}^{\dagger },
                      c_{3 \uparrow}^{\dagger },c_{4 \uparrow}^{\dagger },
                      c_{1 \downarrow}         ,c_{2 \downarrow},
                      c_{3 \downarrow}         ,c_{4 \downarrow} ) \;, \]
while the Weiss field, the cluster Green's function and self-energy
constructed from these operators are $8 \times 8$ matrices.
The exact diagonalization method~\cite{CK:1994} is used
to solve the cluster-bath Hamiltonian Eq.~\ref{eq20} at zero temperature,
which has the advantages of computing dynamical quantities directly
in real frequency
and of treating the large $U$ regime without difficulty. All the results presented here are obtained at quarter filling ($n=1/2$)
far from the particle-hole symmetric case where charge-density-wave
and pairing instabilities coexist.
Qualitatively similar results were found at other densities.

   Fig.~\ref{spin_dblocc_sc.fig}(a) presents the evolution with $U$ of the uniform static cluster spin susceptibility $\chi_{sp}$
and of the double occupancy $n_{d}$ in the NR state
obtained by forcing SC order to vanish.
In the case of the attractive Hubbard model,
$\chi_{sp}$ and $n_{d}$ describe how many fermions turn into local singlet
pairs due to attraction.
In the limit of $U \rightarrow \infty$, $\chi_{sp} \rightarrow 0$ and
$n_{d} \rightarrow n/2$, while
in the opposite limit, $\chi_{sp} \rightarrow 2N(0)$ and
$n_{d} \rightarrow (n/2)^2$.
Here $N(0)$ is the (non-interacting) density of states per spin.
>From weak to strong coupling these thermodynamic quantities are continuous
in CDMFT,
in stark contrast to those of the single site DMFT (see Fig.2 and Fig.3 in
Ref.~\cite{KMS:2001}).
Near $U/t=8$ both $\chi_{sp}$ and $n_{d}$ start to saturate,
indicating that tightly bound bosonic pairs~\cite{MRR:1990}
begin to dominate the physics.

\begin{figure}[tbp]
\includegraphics[width=8.0cm]{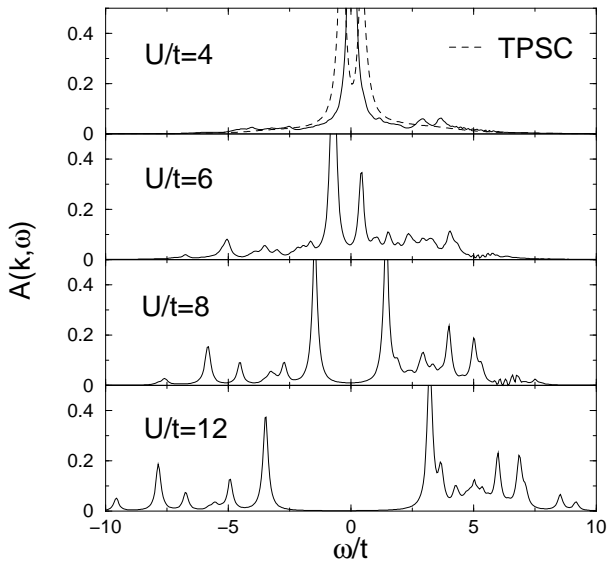}
\caption{Single particle spectral function $A(\vec{k},\omega)$ at $\vec{k}
         = (3/8\pi,3/8\pi)$ near the Fermi surface calculated in the
         NR ground state with a broadening parameter of $0.125t$.
         The dashed curve in the first panel is
         computed from the TPSC at finite temperature of $\beta=t/T=8$ on a
         $64 \times 64$ cluster.
        }
\label{A_w_n=0.5_U.fig}
\end{figure}

   Fig.~\ref{A_w_n=0.5_U.fig} shows additional evidence of the
absence of a first order transition when short-range correlations are
treated explicitly.
Unlike in DMFT, where the spectral function $A(\vec{k},\omega)$
has a peak at $\omega=0$ that disappears at a critical coupling $U_{c}/t \simeq 12$~\cite{Note}
to lead to an insulator,
in CDMFT $A(\vec{k},\omega)$ is a minimum at $\omega=0$ already at $U/t=6$ and we claim that the apparent absence of a gap at $U/t=4$ (the first panel)
is an artifact of the finite size of the cluster used.
In the case of weak coupling,
short-range correlations available in a cluster of size
$N_{c}=4$ are not long enough to lead to a gap, as was found in the
repulsive Hubbard model~\cite{MJ:2001}.
%repulsive Hubbard model~\cite{MJ:2001,KLPT:2003}.
When a large $64 \times 64$ lattice is used to capture long-range correlation
effects within the Two-Particle Self-Consistent (TPSC)
theory~\cite{KAT:2001} (dashed curve) valid at weak coupling, a gaplike
feature does exist even at a finite temperature of $T/t = 1/8$.
QMC calculations on a $16 \times 16$ lattice~\cite{Singer:1999} show the same
result.
Hence, there is enough evidence that
as soon as finite-range spatial correlations are included explicitly,
the first order transition from a Fermi liquid to a non-Fermi
liquid state found in the NR state solution of the DMFT equations
disappears immediately~\cite{Sawatzky:2005},
making the NR state crossover smooth ~\cite{NoteConv}, just as in the SC state.
Physically, the reason why DMFT leads to a first-order transition
is because residual hopping between preformed pairs is of
order $t^{2}/U \times 1/d$ with $d$ the dimension, hence they
have vanishing kinetic energy contribution in large dimension and they localize.
Within CDMFT boson hopping is restored (within the clusters).
With increasing $U$, $A(\vec{k},\omega)$ has a large incoherent spectrum
at high frequencies together with a sharp peak near the gap edge.
A close inspection shows a continuous evolution of high energy peaks
for $U/t \geq 6$.

   The smooth crossover can also be seen by considering Fig.~\ref{spin_dblocc_sc.fig}(b), which shows the imaginary part of the $s-$wave cluster pair correlation
function computed in the NR state for several couplings.
With increasing $U$ beyond $6t$ the peak intensity increases,
while the peak position decreases, scaling as $J$.
This is analogous to the $t^{2}/U$ scaling of the peak position in the spin-spin
correlation function at $\vec{q}=(\pi,\pi)$ in the half-filled Hubbard
model~\cite{KT:2005}.
%In fact, under the particle-hole transformation
%of $\uparrow$ spin electrons on a bipartite lattice
%\[ c_{i \uparrow} \rightarrow (-1)^{i}c_{i \uparrow}^{\dagger} \;, \]
%leaving the $\downarrow$ spins unchanged,
%the attractive Hubbard model is mapped onto the repulsive Hubbard model
%at half-filling with a finite average magnetization of $m=1-n$.
%Under the same transformation the transverse spin-spin correlation
%function at $\vec{q}=(\pi,\pi)$ becomes the $s-$wave pairing correlation
%function at $\vec{q}=(0,0)$.
The gradual change of the pair
correlation function from weak to strong coupling is consistent with the smooth crossover of $A(\vec{k},\omega)$
as a function of $U$.
\begin{figure}[tbp]
\includegraphics[width=8.0cm]{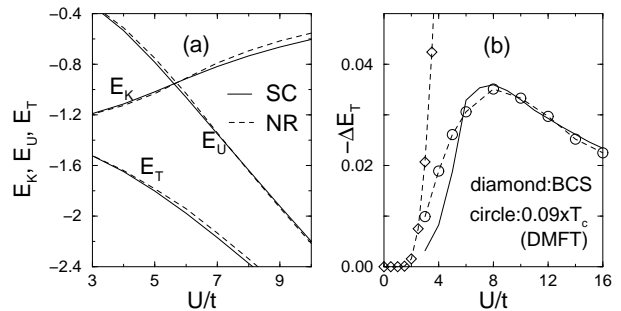}
\caption{(a) Kinetic $E_{K}$, potential $E_{U}$ and total $E_{T}$
         energies computed in the SC and NR ground states
         denoted as the solid and dashed curves, respectively.
         (b) Total energy difference between the SC and NR states
         $\Delta E_{T}$.
         The condensation energy obtained from the BCS theory is shown as
         diamonds. The circles represent $T_{c}$ obtained by Keller
         {\it et al.}~\cite{KMS:2001} within DMFT but multiplied
         by 0.09.
        }
\label{condendation.fig}
\end{figure}

   Next we study the BCS-BEC crossover in the SC state and its
consequences.
Fig.~\ref{condendation.fig}(a) shows the
kinetic $E_{K}$, potential $E_{U}$ and total $E_{T}$
energies computed in the SC and NR ground states
denoted as the solid and dashed curves, respectively.
Generally the energy differences are tiny (only about $1-2\%$), which
implies that the two ground states are energetically very close.
At weak coupling one finds, as in the BCS theory~\cite{BCS:1957}, that while kinetic energy is increased in the SC state
by the broadening of the Fermi surface,
the decrease in the potential energy overcompensates it.
At strong coupling, however, the roles are interchanged.
On a lattice, phase coherence involves breaking up of more local pairs by
virtual hopping (increase in the potential energy) to enhance their mobility
further (decrease in the kinetic energy).
The full inversion of the roles occurs near $U/t=8$, but the decrease in  kinetic
energy happens somewhat earlier ($U/t=6$) in our study.
Although the mechanism for high temperature superconductivity is
debated and the observed gap symmetry ($d-$wave) is different
from the $s-$wave predicted in the attractive Hubbard model,
the current result resembles a recent optical
conductivity experiment in the cuprates~\cite{DSB:2005} where
$\Delta E_{K}$ crosses over from a BCS behavior
($\Delta E_{K} >0$) to an unconventional behavior ($\Delta E_{K} <0$)
as the doping decreases.
In both the SC and the NR states, the kinetic energy scales as $t^{2}/U$.
The difference in the total energy $\Delta E_{T}$ is plotted in
Fig.~\ref{condendation.fig}(b) as a solid curve.
The condensation energy scales with the $T_{c}$ found
by Keller {\it et al.}~\cite{KMS:2001} in DMFT (circles), namely it reaches a maximum near $U/t=8$ and decreases as
$t^{2}/U$ beyond it.
At $U/t=4$ it is already far smaller than that in the BCS theory.
Despite the identical $t^{2}/U$ scaling of $\Delta E_{T}$ in our result and of $T_{c}$ in DMFT at strong coupling,
they differ by a factor of 10 approximatively. This is a upper bound since one expects the mean-field like DMFT estimate for $T_{c}$ to be high because of the neglect of spatial fluctuations.
\begin{figure}[tbp]
\includegraphics[width=8.0cm]{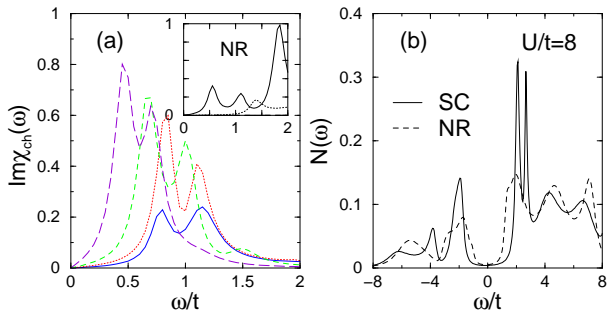}
\caption{(a) Imaginary part of the collective cluster charge excitations
         $\chi_{ch}(\omega)$ at $\vec{q}=0$ in the SC and NR (inset)
         ground states.
         The solid, dotted, dashed and long-dashed curves correspond to
         $U/t=$4, 6, 8 and 12, respectively.
         $\chi_{ch}(\omega)$ for $U/t=$8 and 12 in the inset are invisibly
         small on this scale.
         (b) Density of states $N(\omega)$ computed in the SC (solid) and
         NR (dashed) states at $U/t=8$. A broadening parameter of $0.125t$
         is used in (a) and (b).
        }
\label{bamode_N_w.fig}
\end{figure}

   As was shown above, phase coherence does not lead to a large gain in energy.
The difference between the NR and SC states should manifest itself most dramatically in correlation functions.
We computed the uniform charge-charge correlation function in the cluster in both
states, as shown in Fig.~\ref{bamode_N_w.fig}(a). With increasing $U$
the mode energy decreases, in sharp contrast
with the NR state spectrum shown in the inset.
For $U/t \geq 6$ in the SC state,
the charge excitation forms a sharp resonance much below twice the gap,
while in the NR state (inset)
the weight moves away from $\omega=0$ with weight around $\omega=U$
(not shown) that corresponds to breaking a pair.
In fact it is a collective Bogoliubov mode~\cite{Bogoliubov_mode}
which is separated from the continuum that comes from the breaking of
Cooper pairs at higher energy (not shown). Because a local pair
becomes well-defined only at strong coupling, the difference between
the two states is most pronounced for $U/t=8$ and beyond. The difference between the two states also appears in the local density of state $N(\omega)$ but is more subtle, as shown in
Fig.~\ref{bamode_N_w.fig}(b). Although a pairing gap without phase coherence already exists in the NR state,
the condensation depletes the remnant spectrum from the gap and
builds a sharp spectrum at the gap edge.

   Note that in our approach the phase fluctuations are short range.
Hence, long-wavelength fluctuations do not appear necessary to
lead to kinetic-energy driven pairing. Also, CDMFT allows one to go
beyond Eliashberg theory. The latter is a "strong coupling"
theory in a different sense than that used
in the present paper. It cannot treat the case where
$U$ is of order the bandwidth or larger, i.e. the case where
the normal state consists of bound fermions.
The present study focuses on a $2 \times 2$ cluster with 8 bath sites
since the next smallest cluster size that corresponds to 0.5 filling would
need unreasonable computer resources.
Nevertheless low (but finite) temperature CDMFT+QMC calculations~\cite{KKT:2005}
lead us to believe in the robustness of our results
except at the weakest coupling.

   The increasing fractional deviation of the DMFT~\cite{GKR:2005}
and the CDMFT order parameter and
the energy gap from their corresponding BCS values with decreasing $U$
may be caused not only by the absence of quantum fluctuations
in the BCS theory, but also by
the fact that clusters (in DMFT and CDMFT) are smaller than the spatial extent of the Cooper pairs in that limit.

   Although short-range spatial correlations incorporated in this theory
remove the first-order transition found in the NR state of DMFT,
large frustration may lead back to a first-order transition even
in CDMFT as was found in the half-filled repulsive Hubbard model.
For that model, we found for $t'/t=-0.717$ ($t'/t=1$ in the BEDT model) that a first-order transition does occur
at $U$ close to the bandwidth~\cite{KT:2005}.
In the attractive Hubbard model, frustration could come from magnetic field for example.

   To summarize, we have studied
the BCS-BEC crossover within the two-dimensional attractive Hubbard model
by using the Cellular Dynamical Mean-Field Theory (CDMFT)
both in the normal and superconducting ground states.
Explicit treatment of
short-range spatial correlations in this theory removes
a first-order transition found in DMFT,
making the normal state crossover smooth without a discontinuity.
For $U$ smaller than the bandwidth, pairing is driven by the potential
energy, while in the opposite case it is driven by the kinetic energy,
resembling a recent optical conductivity experiment
in the cuprates.
The condensation energy has a maximum at $U$ equal to the bandwidth and
scales as $t^{2}/U$ at strong coupling.
Phase coherence leads to the appearance of a collective Bogoliubov mode
in the density-density correlation function and to the sharpening of the gap already present in the normal state
spectral function.

\textit{Note added in proof}: Recently
we became aware of DMFT results by Toschi, Capone,
and Castellani~\cite{TCC:2005}, which also showed potential-energy to
kinetic-energy-driven pairing with increasing $U$ in the single
site DMFT.

   The present work was supported by NSERC (Canada), FQRNT (Qu\'{e}bec), CFI
(Canada), CIAR, and the Tier I Canada Research Chair Program (A.-M.S.T.).
Computations were performed on the Elix2 Beowulf cluster and on the Dell cluster of the RQCHP.
We are grateful to M. Randeria and A. Millis for useful comments and information.
A.G acknowledges discussions with M.Capone and the support of and AC-Nanosciences ``Gaz Quantiques'' (Project Nr.201).

%%%%%%%%%%%%%%%%%%%%%%%%%%%%%%%%%%%%%%%%%%%%%%%%%%%%%%%%%%%%%%%%%%%%%%%%%%%%


\begin{thebibliography}{99}
\bibitem{NoteConv} We checked that the converged value of the bath parameters are
independent of whether the starting guess has a large gap or not. In regions where there are two
single-site DMFT solutions, the converged solution also does not depend on which of the two DMFT solutions is chosen
as initial guess of the CDMFT calculation.
\end{thebibliography}
\end{document}